\documentclass{aa}
\usepackage{graphics,epsfig}
\begin{document} 
\title{ HI in Abell 3128} 
\titlerunning{HI in Abell 3128} 
\author{Jayaram N Chengalur\thanks{chengalur@ncra.tifr.res.in}\inst{1,4},
R. Braun\thanks{rbraun@nfra.nl}\inst{2}, 
M. Wieringa\thanks{mark.wieringa@atnf.csiro.au}\inst{3}} 
\institute{ National Centre for Radio Astrophysics, Post Bag 3, Ganeshkind PO, 
Pune, Maharashtra 411\,007, India, \and 
Netherlands Foundation for Research in Astronomy, P.O. Box 2, 
7990 AA Dwingeloo, The Netherlands \and 
The Australia Telescope National Facility, P O Box 76, Epping, NSW 2121,
Australia, \and
Visiting Scientist, NFRA, P O Bus 2, 7990 AA Dwingeloo, The Netherlands } 

\date{Received mmddyy/ accepted mmddyy} 
\offprints{Jayaram N Chengalur} 
\maketitle 

\begin{abstract}
	We discuss Australia Telescope Compact Array (ATCA) HI 21cm 
data for the galaxy cluster A3128. Our observations are intentionally 
relatively shallow, and a blind search through our data cube yields 
(tentative) detections of only two galaxies, of which one is probably
spurious.
	A3128 is part of the ESO Nearby Abell Cluster Survey (ENACS)
(\cite{katgert96}); redshifts are available for 193 galaxies
in the A3128 region. For 148 of these galaxies the redshifts are such
that the HI emission (if any) would lie within our data cube. We use 
the known redshifts of these galaxies to coadd their spectra and thus
improve our sensitivity to HI emission. The technique is fairly 
successful -- the coadded spectra allow detection of an average mass 
content of $\sim 9\times 10^8$ M$_\odot$, almost an order of magnitude 
lower than for direct detection (by which we mean a $5\sigma$ detection
after smoothing to $90\times 90^{''}$ and 300 km/s resolution) of individual 
objects. 
	By dividing the total galaxy sample into subsamples we
find that the gas content of  late type galaxies that lie outside 
the X-ray emitting core of the cluster is substantially higher than
that of those within the core. The fact that for disk galaxies the average 
gas content is higher for galaxies outside the X-ray emitting 
region as compared to those inside implies that these galaxies are 
not well mixed in the cluster potential. Even outside 
the X-ray emitting region the distribution of gas-rich galaxies 
in the cluster is not uniform, we find that gas-rich galaxies are 
concentrated in the east of the cluster. This is consistent 
with earlier analyses of the kinematics of the galaxies in A3128 
which indicate the presence of subclustering.
	In summary we find that coadding spectra is a powerful tool 
for the study of HI in cluster galaxies, and suggest that this
technique could be applied to substantially increase the redshift
range over which such observations could be carried out.

\end{abstract}

\section{Introduction} 
\label{sec:intro}
	
	There is growing evidence that environment plays an important
role in galaxy evolution. In particular, for cluster galaxies, interaction
with  the inter-cluster medium (ICM) as well as gravitational interaction
with other cluster members appears to play an important role. For example, 
numerical simulations suggest that the effect of multiple distant
galaxy encounters (`galaxy harassment') suffered by a galaxy passing 
through a cluster is sufficient to change the morphology of a dwarf spiral
into a dwarf spheroidal~(\cite{moore96}), as well as to sufficiently thicken
the stellar disks of larger spiral galaxies to make them morphologically 
similar to S0s~(\cite{moore99}). Interaction with the ICM is expected to 
have an equally dramatic effect on the gas content of galaxies, with a 
combination of ram pressure and viscous stripping being sufficient to 
strip an $L_\star$ spiral of essentially all of its gas within 
100~Myear~(\cite{quilis00}).

	There is also considerable observational evidence for the effect
of the cluster environment on galaxy evolution. Spiral galaxies in clusters are
known to be deficient in HI as compared to field galaxies (\cite{haynes84}).
However, the molecular gas content (or at least the luminosity of the 
CO emission) appears to be the same for cluster and field galaxies 
(\cite{kenney89}), suggesting that the gas removal mechanisms are most 
effective for the outer parts of the galaxy disk. Synthesis imaging of
galaxies in the Virgo cluster has shown that galaxies near the cluster
center have systematically smaller gas disks than those further out,
and also that the gas disks are asymmetric and have sharp edges on the
side closer to the cluster center (\cite{cayatte90}). The efficient 
removal of gas also necessarily affects star formation -- cluster galaxies
are found to have suppressed star formation rates as compared to 
field galaxies (\cite{balogh98}).
	
	Galaxy clusters, even at low redshifts, are still accreting 
material. Galaxy groups that are falling in to the cluster for the
first time still have substantial gas content and can hence be easily 
identified from their HI emission (e.g. Coma, \cite{bravo00}). 
Similarly, in the Virgo cluster, the HI deficient galaxies show 
considerable substructure (\cite{solanes00}). Another aspect of the
rapid recent evolution of clusters is that clusters at even modest
redshift ($z < 0.5$) appear to have substantially different properties from
those of local clusters. These clusters have a larger fraction of blue galaxies
(\cite{butcher84}) and also have a larger fraction of spirals and a
smaller fraction of S0s as compared to local clusters (\cite{dressler97}).
The gas content of galaxies in moderate redshift clusters is, however,
unknown, since the HI emission from a typical galaxy at these redshifts
is too faint to detect with existing telescopes in reasonable integration
times.

	At the very lowest redshifts the HI content of clusters can be
studied using single dish telescopes, however the poor angular resolution
of these telescopes makes them unsuitable for studies of clusters even
at redshifts of $\sim 0.1$. At these redshifts however aperture
synthesis observations are fairly efficient, since most of the cluster
galaxies fall within a single primary beam.

	We report here on observations of the $z=0.06$ cluster A3128 made
using the Australia Telescope Compact Array (ATCA). The observations were 
relatively shallow, but we coadd the spectra of the different galaxies to 
improve our detection threshold.
The rest of this paper is organized as follows. The observations are
discussed in Sect.~\ref{sec:obs}, the search for HI emission from
individual galaxies in Sect.~\ref{ssec:indi}, and the coadded
spectra from different subsamples in Sect.~\ref{ssec:ave}. 
Sect.~\ref{sec:dis} contains a discussion of the main results
from our analysis. Throughout the paper we use $H_0 = 75$~km/s/Mpc,
$q_0 = 0.5$, $\Lambda=0$ and the angular diameter distance or luminosity 
distance as appropriate. Also, ``heliocentric velocities'' is used 
throughout this paper to mean the quantity cz where z is the heliocentric 
redshift.

\section{Observations and Data Reduction}
\label{sec:obs}

	The observations were conducted at the ATCA from 17 to 20 November 1996.
All observations were conducted
in the 750A array configuration. In this configuration 5 antennas are
stationed such that they give baselines between 77~m and 735~m, the 6th
antenna  is much more distant and its baselines with the inner 5 antennas
vary from 3015~m to 3750~m. Only the data from the 5 inner ATCA antennas
were used in all the processing described below. Since A3128 is somewhat
extended compared to the ATCA primary beam ($\sim 33^{'}$) the 
observations were conducted in a compact mosaic of four pointing centers. 
The center frequency was $ 1339.0$~MHz ($z \sim 0.06$) for all the 
observations. The bandwidth was 32~MHz ($\sim 7600$~km/s) and there were 
a total of 256 spectral channels, giving a channel spacing of 
$\sim 28$ km/s. 

	The observations were spread over $4$ twelve hour sessions, 
and each pointing center was observed once every session. The total 
integration time per pointing center is given in Table~\ref{tab:obs}. 
The standard AT calibrator 1934-638 was observed once every observing
session to determine the absolute flux scale. Phase and bandpass calibration
were done using the source 0407-658, which was observed once 
every 30m.

\begin{table}
\caption{Observation Log}
\label{tab:obs}
\begin{tabular}{llll}
RA(J2000)  & DEC(J2000)  & $\nu$ (MHz) &$\tau$ (hr) \\
3:31:00.01 & -52:45:21.3 & 1339.0 & 9.8 \\
3:31:00.01 & -52:32:37.7 & 1339.0 & 9.6 \\
3:29:36.20 & -52:32:37.7 & 1339.0 & 9.6 \\
3:29:36.20 & -52:45:21.3 & 1339.0 & 9.6 \\
\end{tabular}
\end{table}

	The data were analyzed using standard tasks from the MIRIAD package.
Continuum was subtracted using a second order polynomial fit to the 
uv~data; channels 30 to 239 were used to make the fit. The data from all 
pointing centers were used (after primary beam correction) to make a single 
naturally weighted mosaiced cube with a pixel size of $15^{''}$ and a resolution of 
$81^{''}\times53^{''}$ which corresponds to about $85\times 56$~kpc. The 
edge channels were discarded while making the cube, only channels 30 to 239 
(which corresponds to a velocity coverage of $\sim 6200$~km/s) were used. 
The rms noise per channel is $\sim 1.3$~mJy with slight variation over 
the region used for analysis. 

\section{Analysis}
\label{sec:anal}

\subsection{HI from Individual Galaxies}
\label{ssec:indi}

	Our observations are relatively shallow, the $5\sigma$ mass
limit corresponding to a Gaussian signal with FWHM 300 km/s is $M_{HI}
\sim 8\times 10^9$~M$_\odot$, comparable to the 
$M^*_{HI} \sim 1\times 10^{10}$~M$_\odot$ of local field galaxies 
(e.g. \cite{zwaan97}, corrected for our adopted $H_0$ of $75$~km/s/Mpc). 
Since our spatial coverage is also limited to only $\sim 1.5$ Abell radii,
and galaxies out to $\sim 2$ Abell radii are substantially HI deficient 
(\cite{solanes00}) it is not very surprising that visual inspection of 
the data cube yielded no obvious detection.

\begin{figure}[h!]
\resizebox{8cm}{!}{\includegraphics{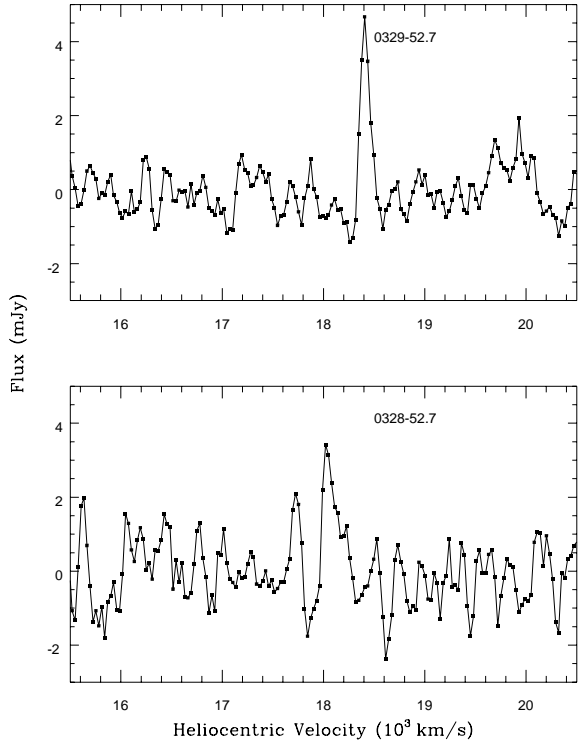}}
\caption{ Spectra of the two locations in the HI cube with emission
greater than the threshold value. The threshold value was chosen such 
that the expected number of spurious detections in the entire cube 
is $\sim 0.5$. The search for emission was ``blind'', i.e over the
entire cube without regard to the location of the cataloged 
galaxies.}
\label{fig:search}
\end{figure}

	A statistically robust blind search was then made over the
entire data cube. Assuming that the distributions of intensities in the
cube is Gaussian, the probability of finding an intensity (purely by
chance) in excess of a given threshold can be computed. The threshold
was set such that the expected number of independent pixels above the
threshold in the entire data cube was $0.5$. The search was done 
(using the AIPS task SAD) at the original spatial resolution, as well 
as with the spatial resolution degraded (via smoothing in the image plane)
to $90^{''}\times 90^{''}$, and for velocity resolutions of 28, 56, 112,
224, and 448~km/s.  Only two `detections' were
found in this blind search; the corresponding spectra are shown in
Fig.~\ref{fig:search}. Table~\ref{tab:search} lists the parameters
derived from these spectra, along with the data for the nearest
cataloged galaxies.  Note that neither of these two galaxies have a
measured optical redshift, so they do not enter into the sample of
galaxies whose spectra we coadd (see Sect.~\ref{ssec:ave}).

	The columns in Table~\ref{tab:search} are as follows. 
Col.~1: Name of the nearest galaxy (IAU Format),
Col.~2: Right Ascension of the nearest galaxy (J2000),
Col.~3: Declination of the nearest galaxy (J2000),
Col.~4: The angular separation ($\alpha, \delta)$, in arc-seconds between
       the nearest cataloged optical galaxy and the HI signal,
Col.~5: Heliocentric velocity of the HI signal,
Col.~6: Peak Flux (mJy) of the HI signal,
Col.~7: Integrated Flux (Jy km/s) of the HI signal, and 
Col.~8: Morphological Type (from \cite{dressler80}).

\begin{table*}
\caption{}
\label{tab:search}
\begin{minipage}[t]{12cm}
\begin{tabular}{lllcllllll}
\noalign{\smallskip} \hline 
Name & RA    & Dec   & $\Delta$(Ra,Dec) & V$_{\rm HI}$ & S$_{\rm peak}$& S$_{\rm int}$& M$_{\rm HI}$     & Type \\
     & J2000 & J2000 & arc-sec          & km/s         & mJy           & Jy km/s      & $10^{9}$M$_\odot$&\\
\noalign{\smallskip} \hline 
0328-52.7 & 03:28:04.1 & -52:44:41.3 & 41 9 & 18021$\pm$ 30 & 3.9$\pm$ 1.3 &0.45$\pm$ 0.1 &6$\pm$ 1.3&...\\
0329-52.7 & 03:29:51.5 & -52:40:44.7 & 4  16& 18407$\pm$ 30 & 5.5$\pm$ 1.3 &0.45$\pm$ 0.1 &6$\pm$ 1.3&S\\
\noalign{\smallskip} \hline 
\end{tabular}
\end{minipage}
\end{table*}

\begin{figure}[h!]
\begin{center}
\resizebox{8cm}{!}{\includegraphics{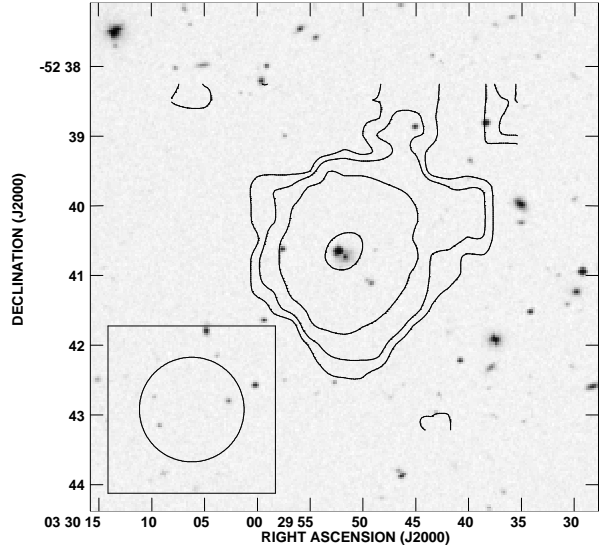}}
\end{center}
\caption{ The HI 21cm contours overlayed over the optical (POSS~II)image
of 0329-52.7. The HI beam ($90^{''}\times 90^{''}$) is shown in the bottom
left corner. The contour levels are 0.5, 0.1, 0.2, and 0.4 Jy/Bm~km/s.}
\label{fig:ovlay}
\end{figure}

Note that our search criteria are generous in that we have ignored 
the uncertainties that would arise from the small deviations of the
noise statistics from a Gaussian distribution. One should hence 
regard these detections as tentative. For 0328-52.7 the HI emission peak is
fairly distant from the optical galaxy, the detection may be spurious.
The HI detection of 0329-52.7 is, however, probably reliable.
Fig.~\ref{fig:ovlay} shows an overlay of the HI data and the optical
data for this galaxy.  This rigorous search for signals 
associated with known galaxies is also useful in interpreting
the results of the next stage of our analysis, namely coadding of 
the spectra of individual galaxies.

\begin{figure*}[t!]
\begin{center}
\resizebox{16cm}{!}{\includegraphics{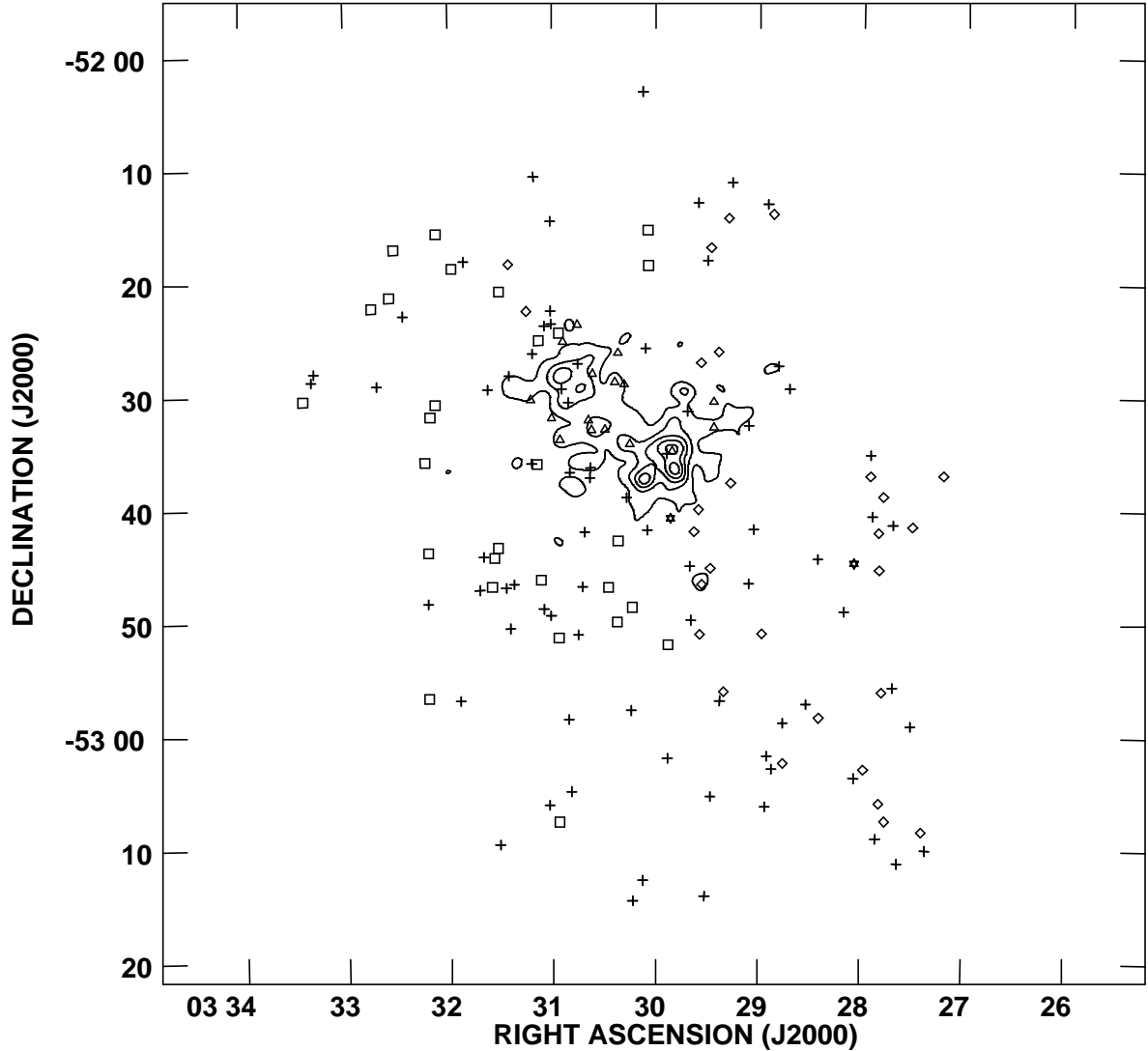}}
\end{center}
\caption{ Position of galaxies in A3128 that have measured redshifts
and lie within our HI data cube. The X-ray emission (from a ROSAT HRI
broadband image) is shown as contours. The two galaxies whose HI 
emission has been tentatively detected (see Sect.~\ref{ssec:indi})
are shown as stars. Neither of these two galaxies has an optically
measured redshift. Late type galaxies are shown as hollow squares,
hollow diamonds and hollow triangles. The hollow squares are ``gas
rich'' on the average, while the hollow diamonds are ``gas poor''
on the average. See Sect.~\ref{ssec:structure} for the definition
of ``gas rich'' and ``gas poor''. Galaxies which are regarded as 
lying within the X-ray emitting region are shown as hollow triangles. 
Crosses are either early type galaxies, or galaxies whose morphological 
type is unknown.}
\label{fig:xray}
\end{figure*}

\subsection{Average HI content }
\label{ssec:ave}

	As discussed in the previous subsection, because our
observations are relatively shallow, we have what are, at best, tentative
detections of two galaxies. A3128 is one of the clusters in the
ESO Nearby Abell Cluster Survey (ENACS), and from multi-fiber spectroscopy,
the redshifts of 193 galaxies in and around A3128 are available 
(\cite{katgert96}, \cite{katgert98}). These redshifts are 
typically accurate to $\sim 50$~km/s or better. It should hence be
possible to considerably improve our sensitivity by coadding the 
spectra of all of these galaxies. Of course, by doing so one is 
restricted to measuring the average HI content of the galaxies 
in the cluster, information on individual galaxies is lost. However,
as we shall show, a judicious choice of subsamples makes coaddition 
a fairly powerful analysis technique.

\begin{figure}[h!]
\vskip -2.75cm
\resizebox{7cm}{!}{\includegraphics{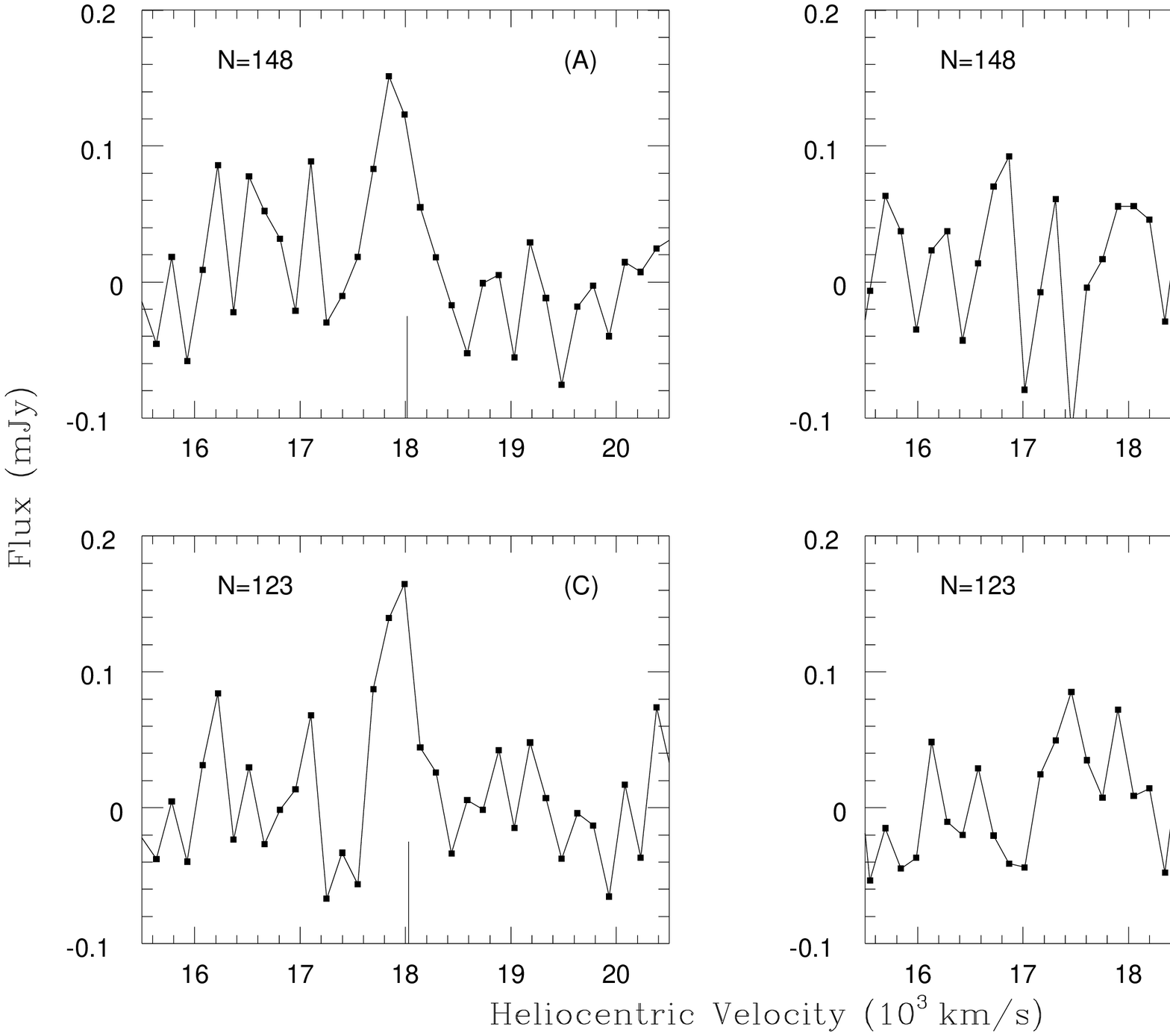}}
\caption{ 
Average and control HI spectra using all ENACS galaxies in
A3128 with measured redshift. 
{\bf (a)} average spectrum over all 148 galaxies. 
{\bf (b)} control spectrum for (a) with randomized velocity shifts.
{\bf (c)} average spectrum over the 123 galaxies outside the cluster core.
{\bf (d)} control spectrum for (c) with randomized velocity shifts. All
spectra have been smoothed to a velocity resolution of $\sim 140$~km/s.
}
\label{fig:all}
\end{figure}

	We start by looking at the average signal from known ENACS
galaxies within the cluster.  The spectra at the location of each
galaxy whose redshift is known were extracted from the $90^{''}\times
90^{''}$ resolution cube, shifted along the velocity axis so that any
HI signal present would appear in the same channel for all spectra, and
then averaged together. In the averaging process each spectrum
is weighted according to its rms (recall that the noise level varies
slightly across our cube, because of the loss in sensitivity at the
edges of the mosaic). The averaged spectrum (all the averaged spectra
shown in this section have been smoothed to a velocity resolution 
of $\sim 140$~km/s) is shown in
Fig.~\ref{fig:all}a. The velocity to which all the spectra have been
shifted is shown by a short vertical line.  Fig.~\ref{fig:all}b shows
another average of these same spectra, the difference being that the
velocity shifts are randomized, i.e. the shift for one galaxy is
randomly applied to some other galaxy. By using the same set of shifts
for both the average spectra we ensure that the statistics of the
shifts applied to the coherently added as well as the randomly added
spectra are the same. As can be seen there is a weak signal ( peak S/N~=~3.5)
present at the correct velocity in the coherently averaged spectrum.

	Galaxies near the center of the cluster are expected to have
lower average HI content, both because of the morphology-density
relation (i.e. because earlier morphological types which have
inherently little HI are dominant in the high density core) and also
because the HI deficiency of late type galaxies increases towards the
cluster center (\cite{cayatte90}; \cite{solanes00}). We have
consequently constructed a subsample consisting of only those galaxies
which lie (in projection) outside the X-ray contours shown in
Fig.~\ref{fig:xray}. The coadded signal from this subsample is shown
in Fig.~\ref{fig:all}c, while the randomized average spectrum (i.e. the
one with the same galaxies, but with the velocities scrambled before
averaging) is shown in Fig.~\ref{fig:all}d. Both the strength and
significance (peak S/N~=~4.0) of the signal seen in Fig.~\ref{fig:all}a are
slightly increased by excluding the galaxies within the X-ray contours.

	In principle it might have been possible that the signal 
seen after coadding the spectra came from just one or two bright
galaxies. In this particular case it is unlikely because, as we
discussed in Sect.~\ref{ssec:indi}, we have no clear detection
of any individual galaxy in the sample.  As a further test every 
individual spectrum was clipped at $2.5\sigma$ (where $\sigma$ 
is the rms in the original individual spectra), the clipped and 
non clipped coadded spectra are essentially identical.

\begin{figure}[h!]
\vskip -2.75cm
\resizebox{7cm}{!}{\includegraphics{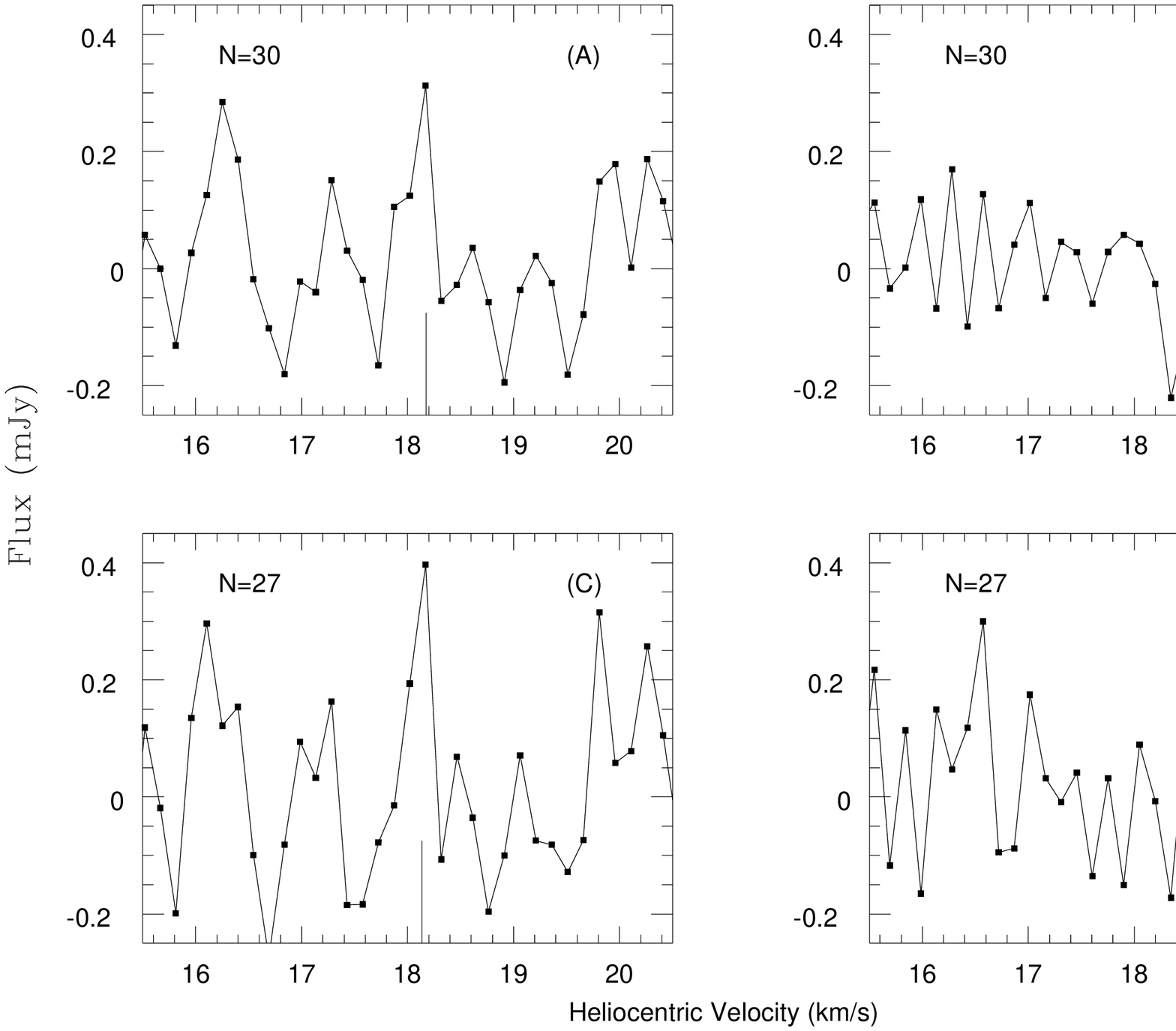}}
\caption{ 
Average and control HI spectra of emission-line ENACS
galaxies in A3128 with measured redshift. 
{\bf (a)} average spectrum over all 30 emission-line galaxies. 
{\bf (b)} control spectrum for (a) with randomized velocity shifts.
{\bf (c)} average spectrum over the 27 emission-line galaxies outside 
         the cluster core.  
{\bf (d)} control spectrum for (c) with randomized velocity shifts. All
spectra have been smoothed to a velocity resolution of $\sim 140$~km/s.} 
\label{fig:em}
\end{figure}

	Having been successful in detecting the averaged HI signal from
all galaxies in the cluster, one could try and determine the averaged
HI signal from appropriately chosen subsamples.  Thirty galaxies in the A3128
sample show optical emission lines. Biviano et al.~(1997) find that the
emission-line galaxies in the ENACS sample are generally spiral
galaxies, i.e. that emission-line galaxies are spirals, but that not
all spirals are emission-line galaxies. The coadded signal for all the
emission-line galaxies in A3128 is shown in Fig.~\ref{fig:em}a, and for
all emission-line galaxies outside the X-ray contours in
Fig.~\ref{fig:em}c.  As before, the comparison randomized average
spectra are shown in Fig.~\ref{fig:em}b and \ref{fig:em}d
respectively. The peak  signal-to-noise ratio in both cases is very low (2.6
and 2.7 respectively) since relatively few galaxies were available. In
any case, it is possible to state that the emission-line galaxies do not
provide a dominant contribution to the average gas mass.

\begin{figure}[h!]
\vskip -2.75cm
\resizebox{7cm}{!}{\includegraphics{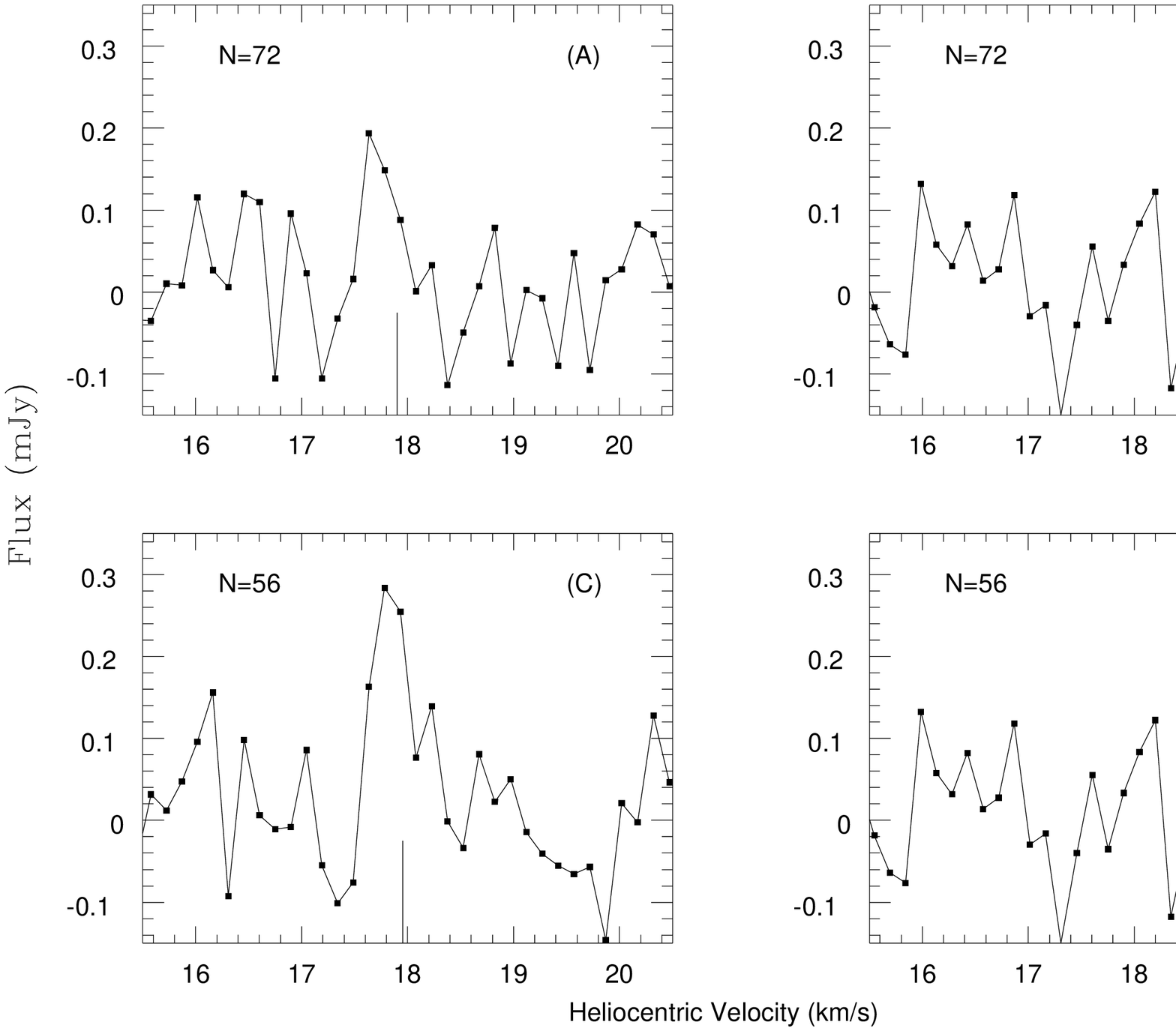}}
\caption{ Average and control HI spectra of ENACS galaxies in A3128 with
morphological type S0 or later and measured redshift. 
{\bf (a)} average spectrum over all 72 late-type galaxies. 
{\bf (b)} control spectrum for (a) with randomized velocity shifts.
{\bf (c)} average spectrum over the 56 late-type galaxies outside the 
          cluster core.  
{\bf (d)} control spectrum for (c) with randomized velocity shifts.  All
spectra have been smoothed to a velocity resolution of $\sim 140$~km/s.}
\label{fig:late}
\end{figure}

	Morphological types (from \cite{dressler80}) are also available
for 130 galaxies in A3128 for which redshift information is also
available from the ENACS survey. Of these 130 galaxies, 108 lie within
our data cube. The coadded signal from all the
galaxies of type S0 and later (where we have deliberately regarded S0s
as ``late-types'' to account for uncertainties in the morphological
typing) is shown in Fig.~\ref{fig:late}a. The coadded signal for the
subset of these galaxies that lie outside the X-ray contours is shown
in Fig.~\ref{fig:late}c. The comparison randomized average spectra are
shown in Fig.~\ref{fig:late}b and \ref{fig:late}d respectively. In
this case a large enhancement in both signal strength and peak
signal-to-noise ratio (changing from 2.8 to 4.0) is seen on constraining
the sample to avoid the cluster core.

	If one restricts the morphological types included in 
the sample to Sa's and later (including types classified 
only as `S') no significant signal is found. There are a 
total of 28 such galaxies (24 outside the X-ray contours). 

\subsection{Substructure}
\label{ssec:structure}

	As we have already seen in Sect.~\ref{ssec:ave}, the galaxies
that lie outside the X-ray contours are more gas-rich on average.
Also, the strongest emission signal is found for galaxies with late 
morphological types and which lie outside the X-ray contours, in line
with what would be expected from theoretical models and
existing HI observations of clusters.

\begin{figure}[h!]
\vskip -2.0cm
\resizebox{5cm}{!}{\includegraphics{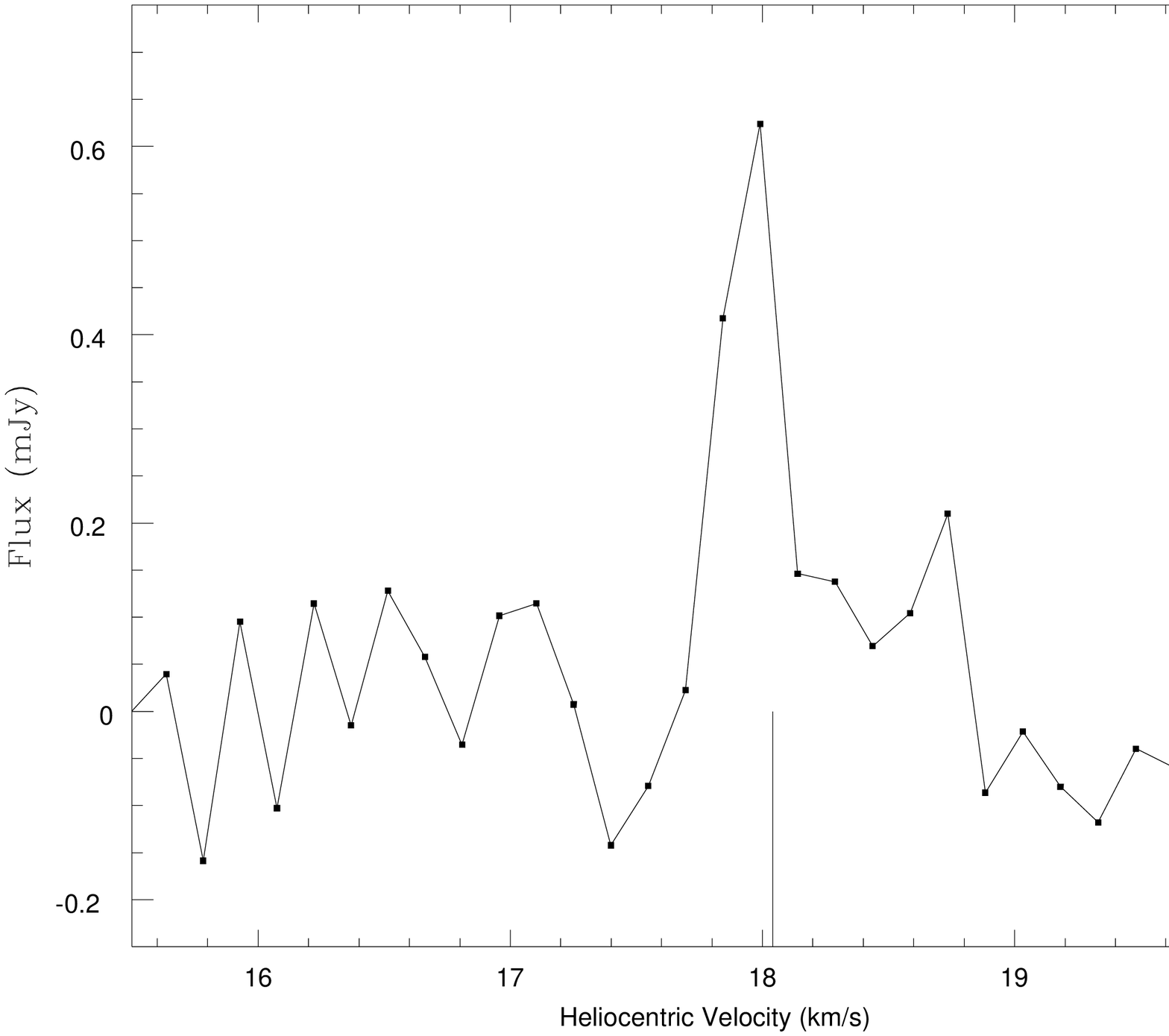}}
\caption{ The coadded spectrum smoothed to 140~km/s velocity resolution 
for the group of 20 galaxies with the 
most significant emission signal. Groups are defined purely on the basis
of proximity in projected separation and independent of HI content of the
individual galaxies. See Sect.~\ref{ssec:structure} for details. }
\label{fig:lpeak}
\end{figure}

	We also tried to examine the distribution of gas-rich
galaxies to check if the average gas content (even for galaxies outside
the X-ray contours) varies with position or not. The procedure we used is
as follows. For each of the 56 late type galaxies outside the X-ray
contours we determined the nearest 20 neighbors (including the target
galaxy itself; by ``nearest'' we mean the 20 galaxies with the smallest
angular distance from the target galaxy). The spectra of this group were
then coadded. For all channels within $\pm 10$ channels (i.e. 
within $\sim \pm 280$~km/s) of the expected
HI signal the ratio of the flux in the channel to the expected rms
noise in the coadded spectrum (i.e. as computed from the rms of the
individual spectra and the number of spectra which contribute to that
channel) was computed. This maximum ``signal to noise ratio'' is
recorded for each galaxy. Target galaxies for which this number is
greater than the median for the entire sample (i.e. ``gas-rich'') are
shown in Fig.~\ref{fig:xray} as hollow squares, and galaxies for
which the number is less than the median (i.e. ``gas-poor'') are shown as
hollow diamonds.  Galaxies which are regarded as lying within
the X-ray emitting region (and omitted from this analysis) are shown 
as hollow triangles. As can be seen, the distribution is far from random, the
``gas-rich'' galaxies are concentrated on the east part of the
cluster. The coadded spectrum with the highest signal to noise ratio is
shown in Fig.~\ref{fig:lpeak}. The HI mass corresponding to this
spectrum is $\sim 2.6\times 10^{9}$~M$_\odot$, and the
$<M{\rm_{HI}}>/<L_R>$ is $\sim 0.1$.  If one uses the entire sample,
instead of only the late type galaxies, one gets essentially the same
result, namely that the galaxies in the east half of the cluster are
more gas-rich on average than those in the west.

\section{Discussion}
\label{sec:dis}

	Abell 3128 is a richness class 3, Bautz-Morgan Type I-II
cluster (\cite{abell89}) with an X-ray luminosity in the $0.5-2$~keV band
of $1.62\times 10^{44}$ ergs/s (\cite{degrandi99}).  Redshifts of 193
galaxies in this field are available from the ENACS survey(\cite{katgert96}).
The HI cube is centered at $\alpha_{2000}=$ 03h30m20s, 
$\delta_{2000}=$ -52d39m15s and is $< 85^{'}$ on a side, which
corresponds to a diameter of $< 3$ Abell radii. 148 of the 193
galaxies for which radial velocities are available lie within our HI
cube.

	The derived quantities corresponding to the spectra in 
Fig.~\ref{fig:all}~to~\ref{fig:late} are summarized in 
Table~\ref{tab:hipar}. The columns in the table are as
follows:
Col.~1: Sample name, i.e. (a)~all the galaxies in the sample, 
        (b)~all galaxies outside the X-ray contours, (c)~all galaxies
        with emission lines, (d)~all galaxies with emission lines 
        and outside the X-ray contours, (e)~all galaxies typed 
        S0 and later, and (f)~all galaxies typed S0 and later which 
        lie outside the X-ray contours,
Col.~2:  Number of galaxies in the sample,
Col.~3:  The HI mass that the spectrum corresponds to. The mass is
        computed from the equation 
       \begin{equation}
	  M_{\rm HI} = 2.35 \times 10^5 D^2 \int S(v) dv
       \end{equation}
       where  $M_{\rm HI}$ is the HI mass in units of solar masses, 
       D is the luminosity distance in Mpc ($\sim 240$~Mpc in this case), the flux
       $S$ is in Jy and the velocity $v$ in km/s.
        The error in the computed HI mass scales as $\sqrt{N_{\rm chan}}\sigma$, where
        $N_{\rm chan}$ is the number of channels over which the signal is spread and $\sigma$
        is the rms noise in the coadded spectrum. For this data set at least, the velocity 
        width of the signal in general increases as the sample size increases. The error in the 
        HI mass hence does not scale as $\sqrt{N_{\rm gal}}$ as might have been 
        expected {\it a priori},
Col.~4: The velocity width of the signal. This is the FWHM of the best
       fit Gaussian,
Col.~5: The average R band luminosity of the galaxies in the sample, and
Col.~6: The ratio of the HI mass to the average luminosity.

\begin{table}
\caption{ Averaged HI content}
\label{tab:hipar}
\begin{tabular}{lrrccc}
\noalign{\smallskip} \hline 
Sample & N & $<M_{\rm HI}>$ &$\Delta$ V &$<L_{R}>$ & 
${<M_{\rm HI}>\over<L_{R}>}$\\
&& $10^8$ M$_\odot$ & (km/s) & $10^{10}$ L$_\odot$ & M$_\odot$/L$_\odot$\\
\noalign{\smallskip} \hline 
All      & $148$  & $ 8.5\pm 2.0$ & $420\pm 47$ & $2.3$ &$ 0.04$ \\
All-out  & $123$  & $ 8.6\pm 2.0$ & $370\pm 47$ & $2.2$ &$ 0.04$\\
Em       & $30$   & $ 8.7\pm 2.6$ & $184\pm 55$ & $1.6$ &$ 0.05$\\
Em-out   & $27$   & $10.4\pm 2.6$ & $179\pm 55$ & $1.7$ &$ 0.06$\\
Late     & $72$   & $ 8.6\pm 2.2$ & $330\pm 45$ & $2.5$ &$ 0.04$\\
Late-out &56     & $16.7\pm 2.6$ & $394\pm 45$ & $2.4$ &$ 0.07$\\
\noalign{\smallskip} \hline 
\end{tabular}
\end{table}

	We note that our sample is not unbiased, since in the ENACS survey
redshifts are generally available only for the brighter galaxies.
Further, the ENACS sample is not a simple magnitude limited sample
because the ease with which the redshift can be measured depends
on the average central surface brightness and there is no simple
relation between the average central surface brightness and the
total magnitude. The ENACS galaxy sample itself (i.e. independent
of whether redshifts were measurable or not) is complete to
$r_{25} \sim 16.5$, which, for a distance of 240~Mpc, corresponds
to an absolute magnitude of $\sim -20.4$. We are hence insensitive
to galaxies fainter than $\sim$ L$^*$. From surveys of field
spirals we know that the gas content (by mass) increases with 
luminosity and that large spirals are the dominant contributors
to the total gas mass in galaxies (\cite{rao93}). In cluster 
environments Valluri \& Jog (1991) had found a tendency for
HI deficiency to increase with increasing optical size, 
however Solanes et al. (2000), using a much larger sample size,
found no such trend. One would hence expect that even in cluster
environments, the bulk of the gas content would be in large spiral
galaxies.

	As was already evident from Fig.~\ref{fig:all} to
\ref{fig:late}, the average gas content of galaxies outside the X-ray
contours is larger than that of those inside (as expected both from the
morphology-density relationship and also from processes in which
interaction with the hot ICM strips the neutral gas from
galaxies). In fact, within the measurement accuracy, our data
are consistent with all the detected gas coming from late type
galaxies outside the X-ray contours. (i.e. the total HI mass in all
the cluster galaxies is equal, within $2\sigma$, to 
the total HI mass in late type galaxies outside the X-ray contours).
However the fact that even for the ``late'' subsample the average
HI content of the galaxies outside the X-ray contours is larger 
than that of those inside the contours suggests that some kind of
gas stripping mechanism due to interaction with the ICM must be 
in operation. It is interesting that the galaxies
``know'' where the X-ray contours are. If the cluster was well mixed,
then the probability that a given galaxy has at some time passed 
through the cluster core would be independent of its current 
position, contrary to what is observed.

	Solanes et al. (1999) have investigated the existence of
substructure in the ENACS sample using diagnostics that are
sensitive to the presence of substructure, but which do not identify
the individual substructures themselves. While all the diagnostics 
suggested the presence of substructure in A3128, the $\Delta$ test 
(\cite{dressler88}) in particular gives a probability of $<10^{-3}$ that
there is no substructure in A3128. Similarly, Biviano et al. (1997)
use the $\Delta$ test to find a probability $<10^{-3}$ that 
there is no substructure in both the entire A3128 galaxy sample, 
as well as the subsample consisting only of non-emission-line 
galaxies. van~Gorkom~(1996) showed that HI content is a good
indicator for substructure, HI rich galaxies are largely confined
to groups that are falling into the cluster for the first time.
Our test for substructure, although necessarily of poorer spatial
resolution, is also indicative of the presence of substructure,
the gas-rich galaxies are confined to the east of the 
cluster. The group of 20 galaxies whose coadded spectrum is shown
in Fig.~\ref{fig:lpeak} have on the average $\sim 2.5$ times more
HI than the remaining late type galaxies outside the X-ray contours.

	Biviano et al. (1997) found that emission-line galaxies are a 
subset of spiral galaxies. De Theije \& Katgert (1999) suggest
that the emission-line galaxies are those spirals which have either
never passed through the core of the cluster or are passing through 
the core of the cluster for the first time. Emission-line galaxies would
then be expected to have a larger HI content than the average spiral
galaxy, however from Table~\ref{tab:search} we find that the average
gas content of the emission-line galaxies and the late type galaxies
is comparable (when one considers the entire subsample). For galaxies
which lie outside the X-ray contours however, late type galaxies are,
on the average, more gas-rich than emission-line galaxies.

	It is worth noting that the average HI mass that we are sensitive
to is $\sim 9\times 10^8$~M$_\odot$, i.e. almost a factor of 10 less
than the $5\sigma$ limit computed in Sect.~\ref{ssec:indi} for 
an individual detection. This makes this technique highly suitable
for extending the redshift range of HI observations of clusters.
In particular, with the WSRT, VLA or GMRT sensitivities, it should be
possible to measure the average HI content of clusters out to redshift
of $\sim 0.5$ in reasonable observing times. This is an extremely interesting
redshift range since from ground based as well as HST observations it 
is now established that there is considerable evolution of cluster 
galaxies between redshifts of 0 and 0.5.  Indeed after submission
of this paper, we learnt that a similar technique had been independently
used by Zwaan (2000) to study the $z=0.2$ cluster A2218. 

	At cosmological redshifts, the number density of Lyman-break
galaxies with redshifts between $3.0 <z < 3.5$ is $\sim 0.4$ 
per square arcmin (\cite{steidel96}). This means that about 
3000 of these galaxies are contained within a single GMRT primary
beam (and a single correlator setting). The excellent GMRT sensitivity
at $z=3.3$ ({\cite{swarup91}) means that coadding the spectra should
allow detection of an average HI mass of $\sim 10^{10}$M$_\odot$ in
an integration time of $\sim 100$~hr. 

\begin{acknowledgements} 
This paper is based on observations with the Australia Telescope 
Compact Array, which is funded by the Commonwealth of Australia for
operation as a National Facility managed by CSIRO. This research 
has made use of data obtained through the High Energy Astrophysics 
Science Archive Research Center Online Service, provided by the 
NASA/Goddard Space Flight Center. This work was partly funded by 
a bezoekersbeurs from NWO to JNC. We are grateful to the referee
(J.~H.~van~Gorkom) for a very careful reading of the paper and numerous
valuable comments, to R. Fanti for assistance with the data analysis,
and to P. Katgert for having supplied the ENACS redshifts in 
advance of publication. Nissim Kanekar's numerous suggestions have
greatly improved the readability of this paper.

\end{acknowledgements}

\end{document}